\documentclass[twocolumn, switch]{article} 

\usepackage{preprint}

\usepackage{amsmath, amsthm, amssymb, amsfonts, symbols}

\usepackage[numbers,sort&compress,square]{natbib}
\usepackage[utf8]{inputenc}	
\usepackage[T1]{fontenc}	
\usepackage{xcolor}		
\usepackage[colorlinks = true,
            linkcolor = purple,
            urlcolor  = blue,
            citecolor = cyan,
            anchorcolor = black]{hyperref}	
\usepackage{booktabs} 		
\usepackage{nicefrac}		
\usepackage{microtype}		
\usepackage{lineno}		
\usepackage{float}			

\usepackage{lipsum}		

\usepackage{newfloat}
\DeclareFloatingEnvironment[name={Supplementary Figure}]{suppfigure}
\usepackage{sidecap}
\sidecaptionvpos{figure}{c}

\usepackage{titlesec}
\titlespacing\section{0pt}{12pt plus 3pt minus 3pt}{1pt plus 1pt minus 1pt}
\titlespacing\subsection{0pt}{10pt plus 3pt minus 3pt}{1pt plus 1pt minus 1pt}
\titlespacing\subsubsection{0pt}{8pt plus 3pt minus 3pt}{1pt plus 1pt minus 1pt}

\title{Communication over Continuous Quantum Secure Dialogue using Einstein-Podolsky-Rosen States}

\usepackage{authblk}

\author[1,*]{Shaokai Lin}
\author[1,*]{Zichuan Wang}
\author[2]{Lior Horesh}
\affil[1]{Department of Computer Science, Columbia University, New York, NY, 10027, USA}
\affil[2]{Mathematics of AI, IBM T J Watson Research Center, Yorktown Heights, NY, 10598, USA}

\begin{document}

\twocolumn[ 
  \begin{@twocolumnfalse} 
  
\maketitle

\begin{abstract}
With the emergence of quantum computing and quantum networks, many communication protocols that take advantage of the unique properties of quantum mechanics to achieve a secure bidirectional exchange of information, have been proposed. In this study, we propose a new quantum communication protocol, called  Continuous Quantum Secure Dialogue (CQSD), that allows two parties to continuously exchange messages without halting while ensuring the privacy of the conversation. Compared to existing protocols, CQSD improves the efficiency of quantum communication. In addition, we offer an implementation of the CQSD protocol using the Qiskit framework. Finally, we conduct a security analysis of the CQSD protocol in the context of several common forms of attack. 
\end{abstract}
\vspace{0.35cm}

  \end{@twocolumnfalse} 
] 


\newcommand\cfootnote[1]{%
  \begingroup
  \renewcommand\thefootnote{}\footnote{#1}%
  \addtocounter{footnote}{-1}%
  \endgroup
}
\cfootnote{Correspondence should be addressed to: sl4299@columbia.edu}

\section{Introduction}
In today’s world where electronic transmission of data plays a vital role in all kinds of communications, its security has become one of the most pressing concerns. Although classical encryption algorithms have been in use and many of them are proven difficult to crack, the development of quantum computing and quantum cryptography \cite{chuang_simple_1995, kielpinski_architecture_2002, gisin_quantum_2002, pirandola_high-rate_2015}, including Shor’s algorithm \cite{shor_algorithms_1994}, have made these classical encryption algorithms no longer impregnable. For example, the RSA algorithm has been traditionally considered difficult to break, but as Ekert showed in 1996, the presumed intractability of factoring large numbers, which algorithms like RSA depend upon, is threatened by the rise of quantum algorithms \cite{ekert_quantum_1996}. Thus, in the quantum era, we need efficient communication protocols that remain robust against both classical and quantum attacks. Fortunately, quantum cryptography promises a new level of security that can be achieved thanks to the unique principles of quantum mechanics. As one of the earliest quantum communication protocols, Quantum Key Distribution (QKD) focuses on distributing a random public key securely to the two communicating parties who would then use the key as a one-time pad to encrypt their messages and transmit them through a classical channel. The most notable QKD protocols are BB84 (the original QKD protocol by Bennett and Brassard) \cite{bennett_quantum_1984} and E91 \cite{ekert_quantum_1991}. Today, QKD has been one of the most studied protocols and has been deployed in certain industrial settings \cite{bunandar_metropolitan_2018, lo_measurement-device-independent_2012, bauml_linear_2018, ma_erratum:_2019}. With the study of quantum networks, the practicality of quantum communication is being scrutinized and further improved \cite{kimble_quantum_2008, pirandola_physics:_2016, frohlich_quantum_2016}.

\paragraph{}
Since then, as an alternative to QKD, Quantum Secure Direct Communication (QSDC) has been put forward and actively studied  \cite{bostrom_deterministic_2002,chuan_secure_2006,deng_secure_2004,dong_quantum_2009,deng_two-step_2003,wang_quantum_2005,yin_efficient_2013,wang_three-party_2007,gao_quantum_2004}. Despite sharing a general goal with QKD, QSDC does not rely on classical channel to transmit encrypted information but rather transmit the encrypted messages through the quantum channel directly. Several QSDC protocols exploit Einstein-Podolsky-Rosen (EPR) pairs \cite{einstein_can_1935} to prevent information leakage \cite{yin_efficient_2013,wang_three-party_2007,gao_quantum_2004}. In 2003 Deng et al. demonstrated the QSDC protocol \cite{deng_two-step_2003} and later Wang et al. presented a similar protocol with improved capacity using high-dimension superdense coding \cite{wang_quantum_2005}.

\paragraph{}
Based on QSDC, a new category of quantum communication protocol, Quantum Secure Direct Dialogue (QSDD), is recently proposed \cite{zheng_quantum_2014}. QSDC and QSDD shares many common features, such as transmitting information through the quantum channel solely \cite{ye_quantum_2015}, but the latter enables users to communicate bidirectionally, which is of practical importance. Until now, two QSDD protocols have been introduced  \cite{zheng_quantum_2014,ye_quantum_2015}, but they both have one crucial disadvantage: the communication over QSDD has to halt after one cycle of information exchange due to the depletion of EPR pairs. In order to continue the exchange, a new handshake between two parties has to be performed to re-establish a secure communication channel. This redundancy of security checks thus creates an unnecessary overhead. By contrast, our proposed protocol, CQSD, has no such inefficiency. In CQSD, every message sent not only transmits information, but also reserves capacity of the next message. Therefore, the two parties can exchange information without interruptions until one of the parties actively closes the channel. Since CQSD eliminates the overhead of redundant initialization, it allows for a "continuous" dialogue. Furthermore, we have shown that CQSD is safe when facing three general forms of attack.

\section{Theoretical Preparation}
\label{Background}
The Bell states are four specific quantum states of two qubits being maximally entangled, as shown below:
\begin{align}
\ket{\Phi^+} = \tfrac{1}{\sqrt{2}} \left( \ket{00} + \ket{11} \right) \\
\ket{\Phi^-} = \tfrac{1}{\sqrt{2}} \left( \ket{00} - \ket{11} \right) \\
\ket{\Psi^+} = \tfrac{1}{\sqrt{2}} \left( \ket{01} + \ket{10} \right)\\
\ket{\Psi^-} = \tfrac{1}{\sqrt{2}} \left( \ket{01} - \ket{10} \right)
\end{align}
A maximally entangled qubit pair is also known as an EPR pair. Through quantum entanglement of the Bell states, we can conduct a Bell inequality check to detect the presence of eavesdroppers or any systematic malfunction in a quantum communication channel that can compromise the security and the reliability of the channel.

\paragraph{}
Now we define the following operators:
\begin{align}
U_1 = \ket{0}\bra{0} + \ket{1}\bra{1} \\
U_2 = \ket{0}\bra{1} + \ket{1}\bra{0} \\
U_3 = \ket{0}\bra{0} - \ket{1}\bra{1} \\
U_4 = \ket{0}\bra{1} - \ket{1}\bra{0}
\end{align}
Note that these operations correspond to certain quantum gates in a quantum circuit implementation. Every operation performed on one qubit of an EPR pair encodes 2 bits of classical information. When the encoded qubit is sent to the other user holding the other qubit of the EPR pair, the recipient can then recover the information from the pair. See the table below for the operator to use given bit strings to be encoded:\\

\begin{table}[h]
 \caption{Operator, corresponding gate, and message}
  \centering
  \begin{tabular}{lll}
    \toprule
    Operator  & Quantum Gate & Message     \\
    \midrule
    $U_1$     & $I$  & 11  \\
    $U_2$     & $X$  & 10  \\
    $U_3$     & $Z$  & 01  \\
    $U_4$     & $ZX$ & 00  \\
    \bottomrule \\
    
  \end{tabular}
  \label{tab:table}
\end{table}

\section{The Continuous Quantum Secure Dialogue (CQSD) Protocol}
\label{CQSD Protocol}
In the CQSD protocol, two communicators, Alice and Bob, intend to communicate with each other directly over a quantum channel, whereas an eavesdropper, Eve, is trying to eavesdrop on this dialogue. Prior to the conversation, Alice and Bob would need to agree on the relationship between the operators ($U_1$, $U_2$, $U_3$, $U_4$) and the corresponding two-bit classical messages. In addition, Alice and Bob should agree on the party initiating the dialogue.  

\paragraph{}
The CQSD protocol begins with the initialization of a secure quantum channel, which requires the initiator of the conversation to prepare and send over qubits to the other party. After a secure channel has been established between Alice and Bob, whenever one party intends to communicate with the other, that party will send three groups of qubits: 1) the qubits that carry the message to be transmitted, 2) the qubits that are used to conduct an eavesdropping check for the current transmission, and 3) the new Bell states to be used in the next exchange. The preservation of readily available EPR pairs is the key to allow for continuous exchange. Here we introduce the steps of the CQSD protocol in detail. 

\subsection*{Step 1. Initial Preparation}
\begin{figure}
  \centering
  \includegraphics[width=0.45\textwidth]{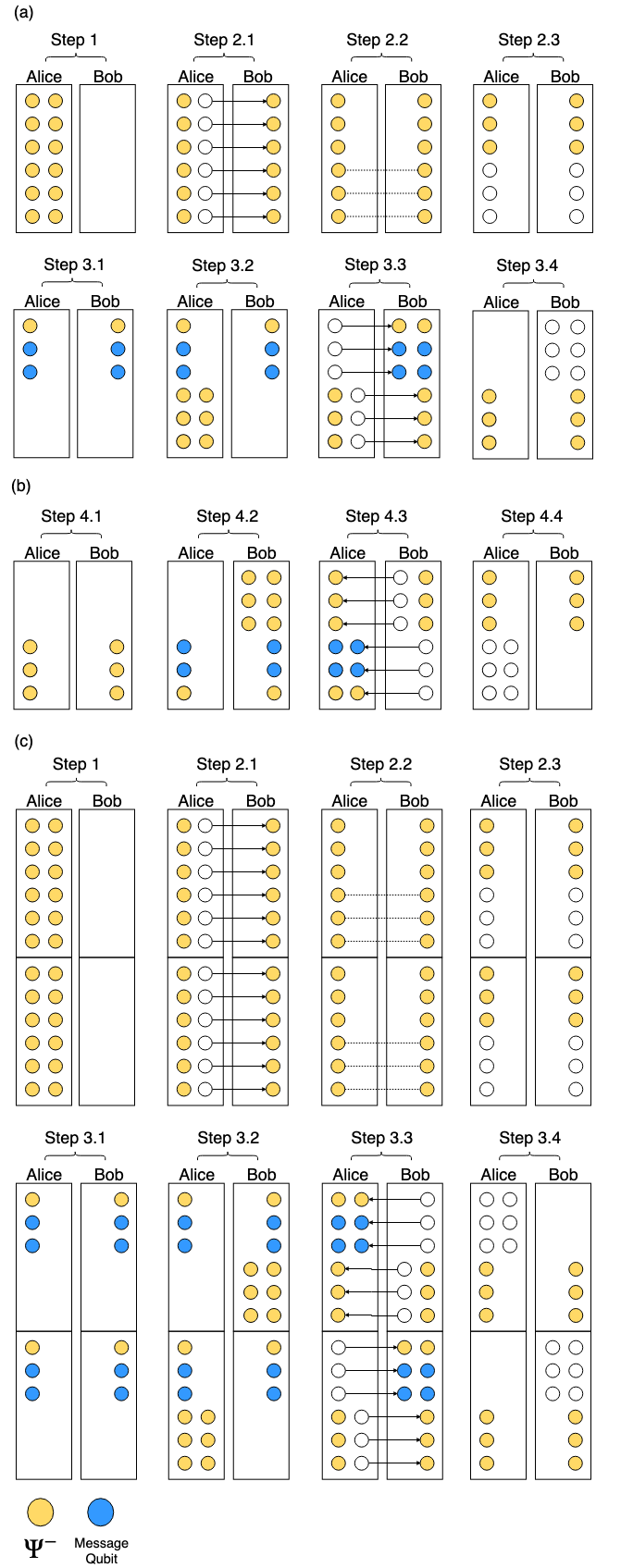}
  \caption{Schematic illustration of CQSD. (a) Alice initiates the communication channel and sends information to Bob. (b) Bob sends information back to Alice. (c) Alice and Bob transmit information to each other simultaneously. Note that $\Psi^-$ is the initialized state of a qubit and the Bell state used for eavesdropper checks. Message qubit indicates the qubit state with the message encoded according to the encoding operators.}
  \label{fig:figure1}
\end{figure}

Alice (or Bob, whomever the two parties designate as the initiator) prepares $2(m+n)$ Bell-state pairs, where $m$ is the number of message qubit pairs and $n$ is the number of “eavesdropper check” qubit pairs that can be sent in one transmission. The choice of $m$ and $n$ can be based on the physical capacity of the device used. After the decision is made by either Alice or Bob, it can be publicly announced. The same $m$ and $n$ will be applied during the entire communication. Similar to QSDD, Alice (and later Bob) prepares all the EPR pairs as $\ket{\Psi^-}$ \cite{zheng_quantum_2014}. Both Alice and Bob will keep using that state to create EPR pairs during the conversation until the conversation stops. From each EPR pair, Alice sends one of the entangled qubits to Bob.

\subsection*{Step 2. Establishing a Secure Channel}
After Bob has received $2(m+n)$ halves of the EPR pairs from Alice, Alice randomly selects $m+n$ particles from her $2(m+n)$ halves as initial check pairs and notify the other party of their positions. Note that the initial security check uses more qubits than subsequent checks. Alice then publicly announces the positions of the check pairs as well as a measuring basis for each pair. Alice and Bob then perform measurements on all of their check qubits in the previously specified bases. After obtaining all the measurements, Alice or Bob can publicly compare the measurements, from which they can measure the error rate of the check pairs. If the quantum communication channel is devoid of an eavesdropper, then Alice and Bob will obtain perfectly anti-correlated measurements and can continue onto the next step. Whereas if there is a third party Eve eavesdropping on the communication, the measurements obtained by Alice and Bob will be disturbed and no longer perfectly anti-correlated. In that case, Alice and Bob terminate the communication directly, forestalling any potential information leakage.

\subsection*{Step 3. Alice Speaks to Bob}
So far, on either Alice’s or Bob’s computer, there are exactly $m + n$ halves of EPR pairs remaining. Alice can then select $n$ qubits at random positions as the check qubits for a specific transmission. She then selects $m$ qubits to encode her message using the unitary operations specified in Table 1. During the encoding process, Alice prepares two groups of new EPR pairs: the first group has the $n$ qubits and the second group has $m$ qubits. If Alice is only using one qubit for the eavesdropper test and one qubit for sending the message, in this scheme, she needs to send four qubits total: one of them being the message qubit, one of them being the eavesdropper test qubit, and two of them being the two entangled qubits from the two new EPR pairs. Alice then sends the qubits over to Bob. After Bob receives the qubits, Alice announces the positions of the eavesdropper test qubits. Subsequently, Bob can perform Bell state measurements on the eavesdropper test qubit pairs and see if the measurements are perfectly anti-correlated in randomly selected bases. If so, the previous transaction is secure. If the results are not perfectly opposite, Alice and Bob would stop the communication immediately, preventing the potential leakage of information. Being sure the communication is secure, he then can perform Bell measurements to read out the message encoded. After obtaining the message, Bob can store the rest of $m+n$ halves of EPR pairs for the next information exchange. For both Alice and Bob, the number of available entangled EPR pairs is unchanged, so that both parties are immediately ready for the next transmission.

\subsection*{Step 4. Bob Speaks to Alice}
Due to CQSD’s nature of preserving the number of entangled EPR pairs after each transmission, Bob can immediately send messages back to Alice without re-initializing the communication channel. So far, on either Alice’s or Bob’s computer, there are exactly $m+n$ halves of EPR pairs remaining. Bob can select $n$ qubits at random positions as the check qubits and select the remaining $m$ qubits to encode his message using the unitary operations specified in Table 1. During the encoding process, Bob also prepares two groups of new EPR pairs: the first group has the $n$ qubits and the second group has $m$ qubits. Bob then sends the qubits over to Alice. After Alice has received the qubits, Bob similarly announces the position of the eavesdropper test qubit. Alice then performs Bell state measurement on the eavesdropper test qubit pairs in randomly selected bases. The check pairs should yield anti-correlated results unless an eavesdropper Eve is present. If Alice and Bob detect Eve, they would again  stop the communication and thereby prevent any information leakage. Otherwise, Alice can decode the entangled message qubits and retrieve the information. Now Alice and Bob again both have $m+n$ halves of EPR pairs. This continuous cycle proceeds until one party decides to terminate the dialogue.

\section{CQSD Implementation and Simulation}
\label{CQSD implementation and simulation}
\begin{figure*}
  \centering
  \includegraphics[width=\textwidth]{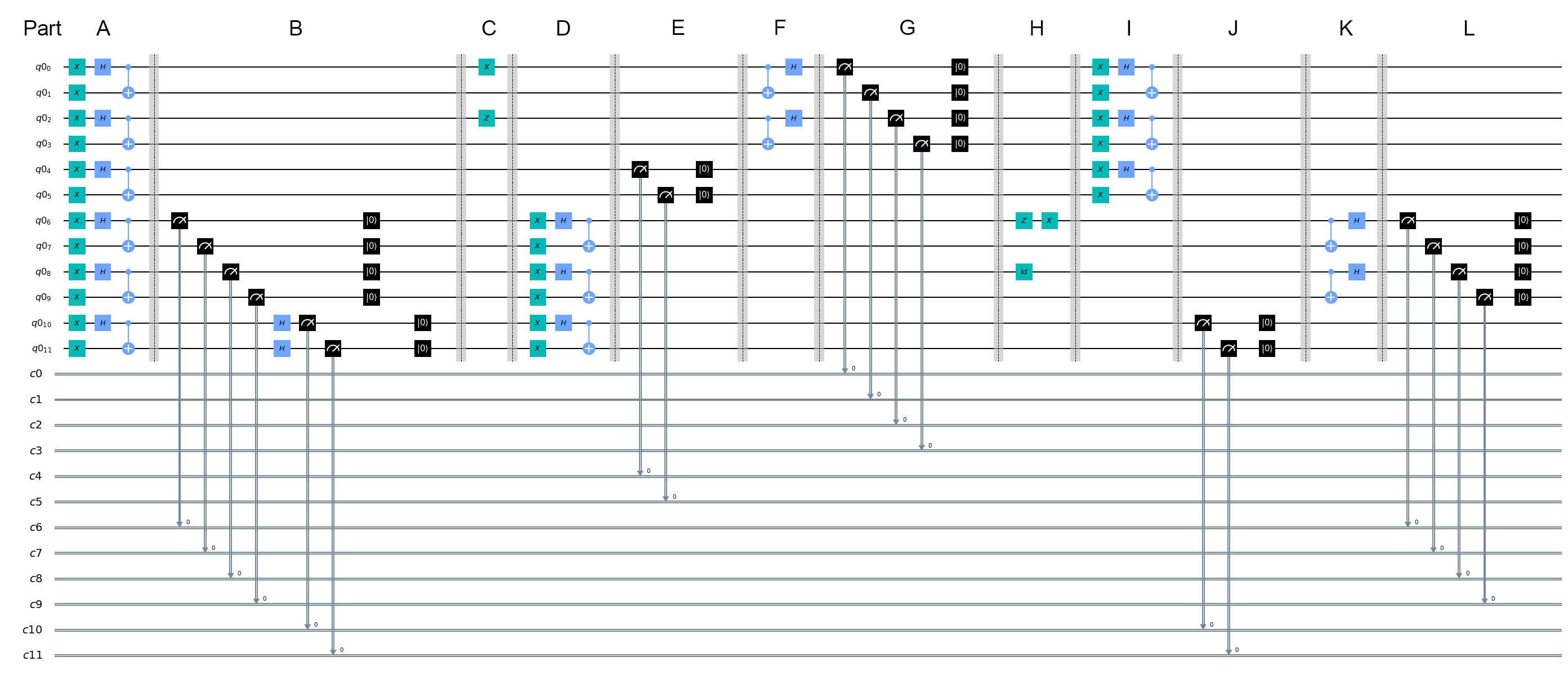}
  \caption{The quantum circuit implementation of the CQSD protocol}
  \label{fig:figure2}
\end{figure*}
We have implemented a simulation of the CQSD protocol using Qiskit. In the simulation, we set up a 12-qubit circuit to simulate the communication between two users, Alice and Bob, each of whom has a 6-qubit quantum computer. The simulation is broken down into three sections: 1. the initial eavesdropper check; 2. Alice speaks to Bob; 3. Bob speaks to Alice. The implementation strictly follows the CQSD protocol specification. Yet, certain details which are currently unable to be implemented are replaced by certain assumptions. One notable assumption is that the transmissions of qubits between users are seamless and perfect. Under this assumption, we are able to split the qubits on the simulation machine into two groups, namely Alice's and Bob's, and directly operate on them.

\paragraph{}
In Figure 2, part A corresponds to step 1 in Figure 1; part B corresponds to step 2.1 to 2.3; part C corresponds to step 3.1; part D corresponds to step 3.2; part E, F and G correspond to step 3.3 and 3.4; part H corresponds to step 4.1 and 4.2; part J, K and L correspond to step 4.3 and 4.4.

\paragraph{}
Because the CQSD implementation is a faithful representation of the CQSD protocol in action, it is apparent that many of the benefits of CQSD are taking effect. For example, the eavesdropper checks at the beginning and during each transmission can effectively detect the presence of an eavesdropper and halt the transmission, preventing any leakage of information. To see this mechanism in action, we reset one of the eavesdropper check qubit in the source code, which is detected by the protocol immediately and triggers the halt. When the disturbance introduced by the eavesdropper is removed, the messages are successfully transmitted and the information exchange can be carried out continuously until the end.

\section{Security Analysis}
\label{Security Analysis}
The CQSD protocol aims to ensure security while enabling continuous communication. Several types of common eavesdropping attacks and the robustness of CQSD against them will be discussed in this section.

\paragraph{}
The first type is called the "intercept-and-resend" attack, in which malicious user Eve intercepts a state being transmitted, retrieves information from the state, and later re-sends the state to the target user. In CQSD, the initial eavesdropper check and the per-transmission eavesdropper check can effectively protect against the intercept-and-resend attack. Assume, without the loss of generality, that Eve is intercepting some of the qubits Alice is trying to send to Bob. In order to perform the attack, Eve measures the qubits in some bases and sends out new ones based on the results. However, the new ones Eve sends to Bob are not entangled with Bob's qubits, so the perfect anti-correlation wouldn't be observed and thus the protocol would alarm Bob. Besides, Eve's measurements of the intercepted qubits would not help when there is an sufficient amount of check qubits as there is no guarantee that she could pick the same measurement bases that would be chosen by Bob later.

\paragraph{}
The second type of attack is the Trojan horse attack, where a malicious eavesdropper probes Alice's apparatus by occupying a part of the quantum channel \cite{gisin_trojan-horse_2006}. Similar to the analysis of preventing this type of attack for QSDD \cite{zheng_quantum_2014}, CQSD can be made safe from the attack using solutions such as installing numerous of photon number splitters (PNS) and filters to detect and exclude invisible photons as well as delayed photons \cite{deng_improving_2005}. The detection of intruding particles can be supported by CQSD's fixed message size. Compared to other communication protocols, which theoretically allow for an arbitrary message size within the device capacity, CQSD imposes a fixed message size to achieve the continuity of message exchange. If the number of incoming particles does not match with CQSD's fixed message size, the device running CQSD can easily identify the presence of intruding particles injected by the malicious user and halt the communication.

\paragraph{}
Man-in-the-middle attack is another attack pattern in which an eavesdropper acts as the facilitator of the communication between Alice and Bob without them knowing, receiving all the messages coming from both parties. Prior research tackles this issue by the distribution of pre-shared secrets, which can be achieved through a variety of techniques \cite{zhang_improving_2007, alleaume_using_2007}.

\section{Conclusions and Future Work}
\label{Conclusions and Future Work}
In this paper, we propose the Continuous Quantum Secure Dialogue (CQSD) protocol, offer an implementation of the protocol, and analyze its security against common types of attack. CQSD demonstrates an improvement over existing protocols by 
offering continuous message flow. Though the protocol design is proposed here, future research is needed to thoroughly assess the CQSD protocol. These research areas include a comparative analysis on the difference between the performance of CQSD and those of other protocols, resource consumption required by CQSD on a quantum machine, implementing CQSD in an experimental setting with photonic devices, as well as generalization of CQSD in higher dimensions. One important area, as we would emphasize for all communication protocols, is their robustness under noisy settings. Analyzing the performance of protocols in quantum networks with different capacities can offer a more comprehensive insight into the actual effectiveness of the protocols \cite{pirandola_end--end_2019}. It is necessary that future studies on CQSD be addressing these various affairs. 

\normalsize
\bibliography{references}

\end{document}